\documentclass[fleqn,11pt]{article}
\usepackage{amsmath,amssymb,amsfonts,graphicx}

\usepackage{amsfonts,amssymb,cite}
\usepackage{graphicx}

\def\onecol{\onecolumn \mathindent 2em}

\def\noi{\noindent}

\newcommand{\Title}[1]{\noi {{\Large\bf #1}}\\[1ex]}

\def\Aunames#1{\noi{\bf #1}}
\def\au#1{${}^{#1}$}
\def\Addresses#1{\medskip\noi \protect
	\begin{description}\itemsep -3pt {\it #1} \end{description}}
\def\adr#1#2{\item[${}^{#1}$]{\it #2}}

\newcommand{\Abstract}[1]{\vskip 2mm \begin{center}
        \parbox{16.4cm}{\small\noi #1} \end{center}\medskip}

\def\email#1#2{\footnotetext[#1]{e-mail: #2}\addtocounter{footnote}{1}}

\def\nqq{\hspace*{-2em}}

\def\qq{\qquad}
\def\cm{\hspace*{1cm}}

\def\wide{\mbox{$\dst\vphantom{\int}$}}

\usepackage{color}

\def\Funding#1{\subsection*{Funding} #1}

\def\Jl#1#2{#1 {\bf #2},\ }

\def\ApJ#1 {\Jl{Astroph. J.}{#1}}
\def\CQG#1 {\Jl{Class. Quantum Grav.}{#1}}
\def\DAN#1 {\Jl{Dokl. AN SSSR}{#1}}
\def\GC#1 {\Jl{Grav. Cosmol.}{#1}}
\def\GRG#1 {\Jl{Gen. Rel. Grav.}{#1}}
\def\IJMPD#1 {\Jl{Int. J. Mod. Phys. D}{#1}}
\def\JETF#1 {\Jl{Zh. Eksp. Teor. Fiz.}{#1}}
\def\JETP#1 {\Jl{Sov. Phys. JETP}{#1}}
\def\JHEP#1 {\Jl{JHEP}{#1}}
\def\JMP#1 {\Jl{J. Math. Phys.}{#1}}
\def\NPB#1 {\Jl{Nucl. Phys. B}{#1}}
\def\NP#1 {\Jl{Nucl. Phys.}{#1}}
\def\PLA#1 {\Jl{Phys. Lett. A}{#1}}
\def\PLB#1 {\Jl{Phys. Lett. B}{#1}}
\def\PRD#1 {\Jl{Phys. Rev. D}{#1}}
\def\PRL#1 {\Jl{Phys. Rev. Lett.}{#1}}

\def\lal{&&\nqq {}}
\def\eq{Eq.\,}
\def\eqs{Eqs.\,}
\def\beq{\begin{equation}}
\def\eeq{\end{equation}}
\def\besub{\begin{subequations}}
\def\esub{\end{subequations}}
\def\bear{\begin{eqnarray}}
\def\bearr{\begin{eqnarray} \lal}
\def\ear{\end{eqnarray}}
\def\earn{\nonumber \end{eqnarray}}

\def\nnn{\nonumber\\ \lal }

\def\yy{\\[5pt] {}}
\def\yyy{\\[5pt] \lal }

\def\dst{\displaystyle}
\def\tst{\textstyle}

\def\e{{\,\rm e}}

\def\im{\mathop{\rm Im}\nolimits}

\def\sign{\mathop{\rm sign}\nolimits}

\def\const{{\rm const}}
\def\eps{\varepsilon}

\def\then{\ \Rightarrow\ }

\newcommand{\vars}[1]{\left\{\begin{array}{ll}#1\end{array}\right.}

\def\eqn#1{\eq\eqref{#1}}
\def\rf{\eqref}

\def\mn{_{\mu\nu}}
\def\MN{^{\mu\nu}}

\def\dpsi{\delta\psi}
\def\da{\delta\alpha}
\def\db{\delta\beta}
\def\df{\delta\phi}
\def\dg{\delta\gamma}

\def\M{{\mathbb M}}
\def\ME{\mbox{$\M_{\rm E}$}}
\def\MJ{\mbox{$\M_{\rm J}$}}
\def\R{{\mathbb R}}

\def\og{{\overline g}}

\def\oR{{\overline R}}
\def\oom{{\overline \omega}}

\def\kappa{\varkappa}

\def\Geff{G_{\rm eff}}
\def\Veff{V_{\rm eff}}
\def\tall{\mbox{$\tst\vphantom{\int^0}$}}
\def\GR{general relativity}
\def\sph{spherically symmetric}
\def\ssph{static, spherically symmetric}

\def\bh{black hole}

\def\wh{wormhole}

\def\asflat{asymptotically flat} 
\def\emag{electromagnetic}
\def\grav{gravitational}
\def\Scw{Schwarz\-schild}
\def\Schr{Schr\"o\-din\-ger}

\usepackage{color}

\begin{document}
\thispagestyle{empty}
\onecol

\bigskip

\Title {On the stability of spherically symmetric space-times\yy in scalar-tensor gravity}
	
\Aunames{Kirill A. Bronnikov,\au{a,b,c,1} Sergei V. Bolokhov,\au{b,2} Milena V. Skvortsova,\au{b,3}\\
			Kodir Badalov,\au{d} and Rustam Ibadov\au{d,4}}
	
\Addresses{\small
\adr a	{Center of Gravitation and Fundamental Metrology, VNIIMS, 
	Ozyornaya ul. 46, Moscow 119361, Russia}
\adr b	{Institute of Gravitation and Cosmology, RUDN University, 
		ul. Miklukho-Maklaya 6, Moscow 117198, Russia}
\adr c 	{National Research Nuclear University ``MEPhI'', 
		Kashirskoe sh. 31, Moscow 115409, Russia}
\adr d {Department of Theoretical Physics and Computer Science, Samarkand State University, 
		Samarkand 140104, Uzbekistan}
	}

\Abstract
   {We study the linear stability of vacuum static, spherically symmetric solutions to the gravitational 
   field equations of the Bergmann-Wagoner-Nordtvedt class of scalar-tensor theories (STT) of gravity,
   restricting ourselves to nonphantom theories, massless scalar fields and configurations with positive 
   \Scw\ mass. We consider only small radial (monopole) perturbations as the ones most likely to 
   cause an instability. The problem reduces to the same \Schr-like master equation as is known 
   for perturbations of Fisher's solution of general relativity (GR), but the corresponding boundary 
   conditions that affect the final result of the study depend on the choice of the STT and a 
   particular solution within it. The stability or instability conclusions are obtained for the Brans-Dicke, 
   Barker and Schwinger STT as well as for GR nonminimally coupled to a scalar field with an 
   arbitrary parameter $\xi$.
    }

\email 1 {kb20@yandex.ru} 
\email 2 {boloh@rambler.ru} 
\email 3 {milenas577@mail.ru} 
\email 4 {ibrustam@mail.ru}

\section{Introduction}

  Stability studies are an important part of theoretical physics in general and theory of gravity 
  in particular. In fact, such a study is necessary for any static or stationary solution of the theory 
  in order to make clear whether or not this solution may lead to a viable model of some real object 
  in Nature, not doomed to decay soon after possible formation. Stability studies embrace 
  a great number of astrophysical phenomena, beginning with structure formation in the early
  Universe and ending with Supernova explosions and gravitational collapse. 
  
  A particular direction of interest in these studies concerns solutions of \grav\ field equations 
  containing scalar fields. Their principal feature is that, unlike those of \emag\ and tensor \grav\ 
  fields, scalar field perturbations possess a monopole degree of freedom that most likely can 
  cause an instability of a particular isolated (most naturally, \asflat) field configuration. The reason is 
  as follows: at least for \sph\ systems, the perturbation equations, after separation of variables, 
  lead to master equations like the one-dimensional \Schr\ equation (see \eqn{Schr} below in this 
  paper) with a certain effective potential $\Veff$ and some boundary conditions that characterize 
  physically plausible perturbations. For different multipolarities $\ell$ of the perturbations, $\Veff$ 
  contains a centrifugal barrier which looks like $\ell(\ell+1)/r^2$. On the other hand, in these master
  equations the role of an eigenvalue, played by an energy level in quantum mechanics, is now played 
  by the squared perturbation frequency $\Omega^2$, and obtaining an eigenvalue $\Omega^2 < 0$ 
  then means that perturbations in question may grow exponentially, which means, in turn, an instability 
  of the original background structure. And, in full similarity to quantum mechanics, a positive
  contribution like $\ell(\ell+1)/r^2$ to $\Veff$ always increases the eigenvalues of the master equation. 
  Therefore, if a system under study is unstable, this result will most probably manifest itself 
  in the case $\ell =0$, that is, for monopole perturbations.    
  
  To our knowledge, such a stability study was for the first time carried out in \cite{kb-hod}, where
  it was concluded that Fisher's \ssph\ solution of GR \cite{fisher} with a scalar massless field is 
  unstable under radial perturbations. In the subsequent years, there have been a great number 
  of stability studies on field configurations containing scalar fields (for reviews see, e.g., 
  \cite{kon-zh11, trap17, stab18}),  mostly focused on \bh\ and \wh\ backgrounds. 
  
  In the present paper, we  consider vacuum solutions of scalar-tensor theories (STT) of gravity, 
  directly generalizing Fisher's solution, that somehow escaped the researchers' attention.  
  These theories, belonging to the Bergmann-Wagoner-Nordtvedt class  \cite{STT1, STT2, STT3},
  have been and remain being among the most popular alternatives to GR due to 
  their simplicity and ability to solve many problems of cosmology and astrophysics. Meanwhile, the 
  solutions in question play in these theories the same role as the one belonging to \Scw's solution 
  in GR and thus deserve a separate stability study. Though, all these solutions possess naked 
  singularities, which casts a certain doubt regarding their relevance. We, however, believe that 
  all space-time singularities (including naked ones) must be suppressed in one or other way 
  by quantum gravity effects, but such effects must encompass only tiny regions around the 
  singularities, outside which the classical laws work quite well, including the perturbation dynamics.
  A subtle question is about the boundary conditions to be imposed at such singularities; anyway,
  we do our best to formulate these conditions in such a way that would survive after singularity
  smoothing. 
  
  Thus we consider monopole perturbations of \ssph\ space-times \MJ\ in STT, comprising their 
  Jordan conformal frame, which are conformal to Fisher's space-time \ME\ that comprises their 
  Einstein conformal frame. The corresponding conformal factors depend on the choice of the STT. 
  Thus we restrict our study to STT with a canonical behavior of the scalar field (those with 
  phantom scalars are conformal to the so-called anti-Fisher space-time, and their properties 
  are drastically different from their canonical counterparts \cite{kb73, stab18}). In addition, we 
  leave aside the cases of the so-called conformal continuation \cite{kb73, kb-CC2}, i.e., those where 
  the conformal mapping from the whole manifold \ME\ leads to only a part of \MJ, making it 
  necessary to continue \MJ\ to new regions where, in general, the effective gravitational constant 
  $\Geff$ becomes negative \cite{kb-CC2, br-star07, skvo10}. 
  
  The structure of the paper is as follows. Section 2 briefly describes the Fisher solution of GR and 
  its counterparts in the general STT \rf{S_J} and four its particular examples to be considered.
  Section 3 presents the perturbation equations that are common for all cases under consideration.
  Section 4 is devoted to the boundary conditions for perturbations that depend on the particular   
  STT and the solution parameters. In Section 5 we numerically solve the relevant boundary-value 
  problems to draw conclusion on the stability or instability of the solutions under study. 
  Finally, Section 6 is a conclusion where the main results of the study are presented in a table.
  
\section{Background}

  The general Bergmann-Wagoner-Nordtvedt STT of gravity is described by the action
  \cite{STT1, STT2, STT3}
\bearr   \label{S_J}
             S_{\rm STT} = \frac 1{16\pi} \int \sqrt{-g} d^4 x
		             \Big[f(\phi) R + 2 h(\phi)\phi^{,\alpha} \phi_{,\alpha} 
  			           - 2 U(\phi) + L_m\Big],
\ear    
  where $R$ is the scalar curvature of space-time, $g = \det(g\mn)$, $f, h$, and $U$ 
  are arbitrary functions of the scalar field $\phi$ ($f(\phi)$ describes a nonminimal 
  coupling between $\phi$ and the curvature, and we assume $f(\phi)> 0$), 
  and $L_m$ is the nongravitational matter Lagrangian. 
  This representation of STT is called the Jordan frame, specified in pseudo-Riemannian 
  space-time $\MJ$ with the metric $g\mn$. The well-known conformal mapping 
\beq              \label{map}
		g\mn = \og\mn/f(\phi)
\eeq  
  converts the theory to the so-called Einstein frame, specified in space-time $\ME$ with 
  the metric $\og\mn$, in which the action takes the form inherent to \GR\ with a minimally 
  coupled scalar field $\psi$,
\bearr   \label{S_E}
             S_{\rm STT} = \frac 1{16\pi} \int \sqrt{-\og} d^4 x
		             \Big[\oR + 2 \eps \og\MN \psi_{,\mu} \psi_{,\nu} 
  			           - 2 U(\phi)/f^2(\phi) + L_m/f^2(\phi)\Big],
\ear      
  where the bars mark quantities obtained from or with $\og\mn$, and the fields $\phi$ and
  $\psi$ are related by 
\beq  			\label{phi-psi}
		\frac {d\phi}{d\psi} = \frac{\sqrt 2 f(\phi)}{\sqrt{|D|}}, \qq
		D = fh + \frac 32 \Big(\frac{df}{d\phi}\Big)^2, \qq \eps = \sign D.
\eeq  
  In the case $\eps =1$ the field $\psi$ is of canonical nature, and if $\eps =-1$, it is of 
  phantom nature, i.e., is characterized by negative kinetic energy.
  
  It is clear that if a solution to the field equations is known in $\ME$, its counterpart in 
  $\MJ$ is easily obtained using \rf{map} and  \rf{phi-psi}.

  We will consider \sph\ space-times being solutions to the theory \rf{S_J} under the 
  conditions $L_m =0$ (no matter) and $U(\phi) =0$ (a massless scalar field).
  
  In the general case, a \sph\ metric can be written in the form (see, e..g., \cite{LL})
\bearr         \wide        \label{ds}
     ds^2 = \e^{2\gamma}dt^2 - \e^{2\alpha} du^2 - \e^{2\beta}d\Omega^2,
                                                        \label{ds-sph}
     \ \ \qq   d\Omega^2 = d\theta^2 + \sin^2 \theta\, d\phi^2,
\ear
   where $\alpha,\ \beta,\ \gamma$ are, in the general case, functions of the radial
   coordinate $u$ and the time coordinate $t$. We will also use the notation
   $r \equiv \e^\beta$; thus $r$ is the radius of a coordinate sphere $u=\const,\ t=\const$, 
   or, which is the same, the Schwarzschild radial coordinate. Even in the static case, with
   only $u$ dependence of the metric coefficients, there is a freedom to fix the choice of
   the radial coordinate by specifying a relation between the functions 
   $\alpha,\ \beta,\ \gamma$.
  
  For the system under study, \ssph\ solutions are well known in $\ME$
  \cite{fisher,ber-lei} and can be written in the general form \cite{kb73} in terms of the 
  harmonic radial coordinate $u$ specified by putting $\alpha(u) = 2\beta(u) + \gamma(u)$.
  The solution has the form \cite{kb73, br-book} 
\bearr                                                        \label{fish}
	    ds_E^2 = \e^{-2mu} dt^2 - \frac{\e^{2mu}}{s^2 (k,u)}
             		\biggl[\frac{du^2}{s^2 (k,u)} + d\Omega^2 \biggr];
\nnn
         \psi = C u, 		\qq   	 k^2\sign k = \eps C^2 + m^2
\ear
  (the index $E$ indicates the Einstein frame). The function $s(k, u)$ is defined as 
\beq  				 \label{s(k,u)}
		s(k,u) :=  \vars {              
                    k^{-1} \sinh ku, & k>0;\\
                    u,               & k=0;\\
                    k^{-1}\sin ku,   & k<0.     }
\eeq  
  Without loss of generality, we can assume $u \geq 0$, so that $u = 0$
  corresponds to spatial infinity at which the metric is \asflat, and the constant $m$ 
  has the meaning of \Scw\ mass since $g_{tt}\approx 1 - 2m/r$ at small $u$. The constant 
  $C$ has the meaning of a scalar charge.
  
  For a canonical scalar field, $\eps=1$, we evidently have $k > 0$, while for a phantom 
  field there are three branches corresponding to $k > 0, \ k=0$, and $k < 0$.
   
  In the case $k > 0$, it is easy to pass on to the ``quasiglobal'' coordinate
  \cite{br-book} $x$ defined by the condition $\alpha+\gamma =0$ in the metric \rf{ds}.
  Indeed, the substitution $\e^{-2ku} = 1 - 2k/x$ brings the solution to the form
\bearr                                                       \label{fish-a}
     ds_E^2 = \Big(1 - \frac {2k}{x}\Big)^a dt^2 
     - \Big(1 - \frac {2k}{x}\Big)^{-a} dx^2 - \Big(1 - \frac {2k}{x}\Big)^{1-a}x^2 d\Omega^2,
\nnn                                                      
     \psi = -\frac{C}{2k} \ln \Big(1 - \frac {2k}{x}\Big),
\ear
  with the constants related by
\beq                                                        \label{int-a}
       a = m/k, \cm  a^2 = 1 - \eps C^2/(k^2).
\eeq
  
  In what follows, we will consider the solution \rf{fish-a} for $\eps =+1 \then a <1$
  and its counterparts in the Jordan frame $\MJ$ for a few examples of STT, in particular:
\begin{enumerate}
\item  
	The Brans-Dicke theory \cite{BD-STT}:
\besub	           \label{BD}
\bearr           \label{BDa}
		f(\phi) = \phi, \qq h(\phi) = \frac{\omega}{\phi}, \qq \omega = \const \ne -3/2,
\\ \lal           \label{BDb}
		\then\ \psi = \psi_0  + \frac {\oom}{\sqrt 2} \log |\phi|, \qq \psi_0=\const, \qq 
		\oom = \sqrt{|\omega + 3/2|},
\yyy             \label{BDc}
		ds_J^2 = \e^{-\sqrt 2 (\psi-\psi_0)/\oom} ds_E^2,				
\ear
\esub
   where $ds_J^2$ is the metric in $\MJ$.
\item
	Barker's theory \cite{barker}, in which the effective gravitational constant is really
	a constant:
\besub	           \label{Bark}
\bearr            \label{Bark-a}
		f(\phi) = \phi, \qq h(\phi) = \frac{4-3\phi}{2\phi(\phi-1)},
		\ \then\ \psi = \psi_0 + \arctan \sqrt{\phi -1},
\yyy            \label{Bark-b}
				\phi = \frac 1 {\cos^2 (\psi-\psi_0)},  \qq
				 ds_J^2 = \cos^2 (\psi-\psi_0)\ ds_E^2,	
\ear
\esub	
   where we have assumed $\phi >1$, corresponding to a canonical nature of $\psi$.   
\item
	The theory motivated by Schwinger \cite{schwg} according to \cite{bruk94}: 
\besub	           \label{Swg}
\bearr            \label{Swg-a}			
		f(\phi) = \phi, \qq h(\phi) = \frac{K-3\phi}{2\phi^2}, \qq K = \const
		\ \then\   \psi = \psi_0 + \sqrt{K/\phi},		
\yyy 			  \label{Swg-b}			
		\phi = \frac K {(\psi-\psi_0)^2}, \qq ds_J^2 = \frac {(\psi-\psi_0)^2}{K}\ ds_E^2.	
\ear
\esub	
\item
	Conformal or nonconformal nonminimal coupling:
\beq                \label{nonmin}
		f(\phi) = 1 - \xi \phi^2, \qq  h(\phi) = 1, \qq \xi = \const.	
\eeq	
     This kind of STT splits into four cases, each requiring a separate treatment,
     to be discussed later: (a) $\xi = 1/6$ (conformal coupling), 
     (b) $0 < \xi < 1/6$, (c) $\xi > 1/6$, and (d) $\xi < 0$.  
\end{enumerate}  
  
\paragraph{A note on conformal continuations.} When two space-times are conformally related,
  it is sometimes possible that a singularity in one of them (say, \ME) maps into a regular surface
  in the other, say, \MJ, which should be then continued beyond this surface.
  Thus the whole manifold \ME\ maps to only a part of \MJ, and the stability problem should be 
  formulated for the whole \MJ. Some general properties of such conformal continuations of
  \ssph\ space-times have been studied in \cite{kb-CC1, kb-CC2}. It is clear that in such cases
  the stability problem must be formulated separately from the general case in which the 
  conformal mapping \rf{map} puts the points of \MJ\ and \ME\ in one-to-one correspondence.
  The present study is devoted to this general case, but we will mention some existing results
  for conformally continued space-times.
  
  A conformal continuation due to \rf{map} is only possible if in \ME\ different metric coefficients
  turn to zero or infinity in the same manner, making it possible to be cured by the proper 
  behavior of a conformal factor. In Fisher's solution \rf{fish-a} this happens if $a = 1/2$,
  so that $g_{tt} \sim g_{\theta\theta} \sim \sqrt{x-2k}$ as $x \to 2k$. 
  It is, however, only a necessary condition: to cure the singularity, the conformal factor must 
  have the precisely opposite behavior. Among the enumerated STT, this happens only in the 
  following cases: (i) in the Brans-Dicke theory with $\omega =0$, and (ii) in nonminimal coupling 
  theories with $\xi > 0$.     
      
\section{Perturbation equations}
   
  Let us now consider \sph\ perturbations of the solutions in $\ME$, assuming that 
  there is a certain \ssph\ solution of the form \rf{ds} with $\psi = \psi(u)$ and introduce
  a perturbed unknown function
\[
    \psi(u,t)= \psi(u)+ \dpsi(u,t)
\]
  and similarly for the metric functions $\alpha,\ \beta,\ \gamma$. 
  In this case, there is only one dynamic degree of freedom related to the scalar field
  because gravitational perturbations cannot be \sph\ (monopole). 
  Accordingly, an analysis of the perturbed field equations leads to a single wave 
  equation for $\dpsi$, and its further study must lead to a conclusion on the stability or
  instability of the static configuration. Technically, this single wave equation can be most
  easily obtained using the perturbation gauge $\db \equiv 0$ (which corresponds to choosing 
  a particular reference frame in perturbed nonstatic space-time) and excluding $\da$ and 
  $\dg$ from the field equations, as is described in detail in \cite{stab11,stab18, br-book}, 
  where it is also shown that the resulting wave equation is actually gauge-invariant and
  therefore  describes the behavior of real perturbations of our system rather than possible
   coordinate effects.
  
  In terms of an arbitrary radial coordinate $u$, the wave equation for $\dpsi$ reads
  \cite{stab11, stab18, br-book, kb-hod}
\beq  			\label{wave-eq}
      \e^{2\alpha-2\gamma} \delta\ddot\psi  -\dpsi'' 
      - \dpsi' (\gamma'+ 2\beta'-\alpha') + W(u) \dpsi =0,
      \qq
      W(u) \equiv  - \e^{2\alpha -2\beta}\frac{2\eps \psi'{}^2}{\beta'^2},
\eeq  
  where the dot denotes $\partial/\partial t$, the prime $\partial/\partial u$, 
  and for a canonical scalar field 
  under consideration we have $\eps = 1$. Since the background is 
  static, we can, as is usually done in such problems, separate the variables by putting 
\beq  
 			\dpsi = \e^{i\Omega t} X(u),\qq \Omega = \const,
\eeq  
  then \eq \rf{wave-eq} reduces to an equation for $X(u)$,
\beq  			\label{eq-X}
        X'' + (\gamma'+ 2\beta'-\alpha') X' 
        					+ [\e^{2\alpha-2\gamma}\Omega^2 - W(u)] X =0,
\eeq  
  which can be called the {\bf master equation} for small linear perturbations of our system.
  Explicitly for Fisher's solution \rf{fish} this equation reads
\beq 			\label{eq-Xu}
		X'' - \bigg[ \frac{k^4 \e^{4mu}}{\sinh^4 ku}\,\Omega^2
				\ + \ \frac{2k^2 (k^2 - m^2)}{(k \cosh ku - m \sinh ku)^2}\bigg] X =0. 
\eeq    
  To bring the master equation to a canonical form, we substitute 
\beq 			\label{X_to_Y}
		X(u) = \e^{- \beta} Y(z),
\eeq  
  where $\beta = \log r$ is, as before, the metric coefficient from the static solution; 
  $z$ is the tortoise coordinate defined by the condition $\alpha = \gamma$ in \rf{ds}
  and related to an arbitrary radial coordinate $u$ in \rf{ds} and the quasiglobal
   coordinate $x$  in \rf{fish-a} by, respectively,
\beq                \label{u_to_z}
       \frac {du}{dz} = \e^{\gamma (u)-\alpha(u)}, 
       \qq
       \frac {dx}{dz} = \Big(1 -\frac{2k}{x}\Big)^a. 
\eeq  
  At finding $z$ from \rf{u_to_z}, we choose the integration constant so that 
  $z=0$ corresponds to $x = 2k$.
  We obtain the following \Schr-like equation for $Y(z)$ \cite{kb-hod, stab11}:
\beq                                                        \label{Schr}
      \frac {d^2 Y}{dz^2} + [\Omega^2 - \Veff(z)] Y =0,
\eeq
  with the effective potential expressed in terms of an arbitrary coordinate $u$ as
\beq  			\label{Veff-u}
		\Veff(u) = - \e^{2\gamma-2\beta}\frac{2\psi'{}^2}{\beta'{}^2}
		+ \e^{2\gamma-2\alpha}\Big[\beta'' + \beta'(\beta'+\gamma'-\alpha')\Big],
\eeq  
  or for the solution \rf{fish-a} written in terms of $x$ \cite{stab11}, 
\beq  			\label{Veff-x}
	\Veff(x)=\frac{k\big[2ax^3 - 3(1+a)^2 kx^2 
			+2(3+4a+3a^2+2a^3)k^2x - (1+a)^4k^3\big]}
						{\big[x - k(1+a)\big]^2 x^{2+2a} (x-2k)^{2-2a}}.
\eeq  
  Note that the tortoise coordinate $z$ is expressed in terms of $x$ according to \rf{u_to_z}
  with the aid of the hypergeometric function, and, as a result, \eqn{Schr}
  has quite a complicated form. However, near the singularity $x =2k$ we can write simply 
\beq                 \label{Veff-2k}
		z \approx \frac{(2k)^a}{1-a} (x-2k)^{1-a},\ \  \then \
				\Veff \approx - \frac 1 {4z^2}.
\eeq  
  As a result, the general solution to \eqn{Schr} near $x=2k$, with any $\Omega$, has the
  approximate form
\beq 				\label{Y-2k}
		Y(z) \approx \sqrt z (C_1 + C_2 \log z), \qq C_1, C_2 = \const.
\eeq  
  At large $z \approx x$, the potential behaves as $\Veff \approx  2ak/z^3$.   
  The behavior of $\Veff$ versus $y = x-2k$ and the parameters $a$ and $k$ is illustrated 
  in Fig.\,1.
  
\begin{figure*}[ht]
\centering
\includegraphics[width=8cm]{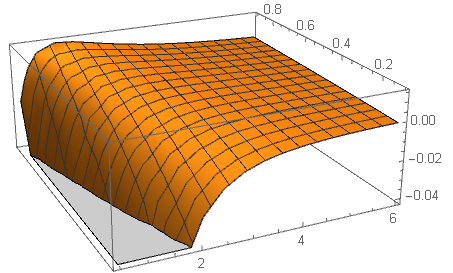}\qq
\includegraphics[width=8cm]{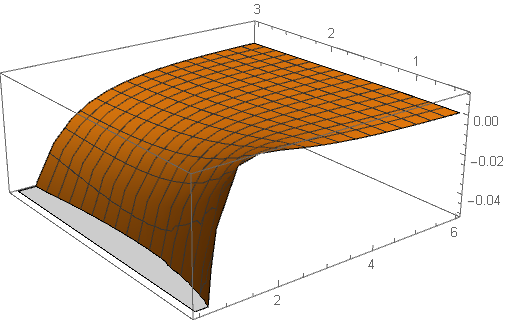}
\caption{The effective potential $\Veff$ as a function of $y = x-2k$ and $a \in (0.1, 0.9)$
	at $k =1$ 	(left panel) and as a function of $y$ and $k \in (0.5, 3)$ at $a = 0.5$ 
     (right panel).}
\end{figure*}  
  
\section{Stability conditions}  
  
  Equation \rf{eq-X} or \rf{Schr} can be used to analyze the stability of background static 
  solutions under \sph\ perturbations. Thus, if there is a nontrivial solution to \rf{eq-X} 
  or \rf{Schr} such that $\im \Omega < 0$, for which some physically meaningful boundary
  conditions hold at the ends of the range of the radial coordinate $u$ or $z$ 
  (including, in particular, absence of ingoing waves), then we can conclude that the 
  background system is unstable because the field perturbation $\dpsi$ 
  can exponentially grow with time. Otherwise the system is stable under this kind of 
  perturbations in the linear approximation.
  
  Since the conformal mapping \rf{map} can be treated as just a substitution in the field
  equations, \eqn{Schr}, written for the Einstein frame can be used for studying the stability 
  of STT solutions in their Jordan frame as well, but the relevant boundary conditions will
  be different for solutions of different STT. In what follows we will discuss these boundary
  conditions and their outcome for the examples of STT enumerated in Section 2.

\subsection{Solutions of particular theories} 
  
\paragraph{Fisher's solution of \GR.} Let us begin with the solution \rf{fish-a}. At spatial 
  infinity $z\sim x\to \infty$ the solution is \asflat, and $\psi \approx C/x \to 0$, therefore, 
  a natural boundary condition for \eqn{Schr}, providing a finite scalar field energy 
  contribution from the perturbations, is $\dpsi/z \to 0 \then Y\to 0$ as $x\to\infty$. 
  
  At the central singularity $x\to 2k$ (corresponding to $z \to 0$), we have $\psi \to \infty$, 
  and, following \cite{kb-hod} and \cite{stab11}, it is reasonable to impose a minimal 
  boundary condition providing applicability of the perturbation scheme, admitting
  divergence of $\dpsi$ but not faster than $\psi$, that is, $|\dpsi/\psi| < \infty$.  
  Thus the admissible behavior of $\dpsi$ and $Y(z)$ as $z\to 0$ will be 
\beq        \label{fish-0} 
		\dpsi \sim \log(x-2k) \sim \log z \ \ \then \ \ Y(z) \sim \sqrt z \log z,
\eeq  
  since $z \sim (x-2k)^{1-a}$, $r \sim  (x-2k)^{(1-a)/2} \sim \sqrt z$, and 
  $Y(z) \sim r \dpsi \sim \sqrt z \dpsi$.
  
  Comparing \rf{fish-0} with \rf{Y-2k}, we see that all solutions to \eqn{Schr} with 
  any $\Omega^2$, including those with $\Omega^2 < 0$, satisfy the boundary 
  condition at $z=0$, and this evidently concerns solutions obeying the condition 
  at $z\to \infty$. In particular, there are physically meaningful perturbations $\dpsi$
  growing with time as $\e^{|\Omega|t}$, which manifest a catastrophic instability of 
  Fisher's solution, in agreement with \cite{kb-hod, stab11}.
  
\paragraph{Solution of the Brans-Dicke theory.}
  The solution has the form \rf{BD}, where $ds^2_E$ is given by \rf{fish-a} for $k >0$.
  The Brans-Dicke scalar field has the form 
\beq                 \label{phi-BD}
			\phi = \phi_0 \e^{-\sqrt 2 \psi/\oom}, \qq \phi_0 = \const.
\eeq  
  In this and other examples of STT, the boundary conditions for perturbations at a possible 
  singularity of the background solution should naturally be formulated in terms of the 
  scalar field $\phi$ involved in the action \rf{S_J},
  but a problem is that there is reparametrization freedom: in \rf{S_J} one may 
  arbitrarily substitute $\phi = \phi(\Phi)$ without changing the physical content of the 
  theory (under reasonable requirements to this substitution). It, however, seems most
  natural and reasonable to unambiguously formulate the boundary conditions for the 
  nonminimal coupling function $f(\phi)$: namely, $|\delta f/f| < \infty$. Thus, if 
  $f(\phi)$ is finite, we require a finite perturbation $\delta f$, and if not, the perturbation
  must not grow faster than $f$. Moreover, if $f \to 0$, it is also reasonable to require
  $\delta f \to 0$, at least because the effective gravitational constant $G_{\rm eff}$ is 
  proportional to $1/f$, and one can hardly admit perturbations growing faster than
  $G_{\rm eff}$ when the latter blows up. 
  
  In examples 1--3 of STT enumerated above, we have $f(\phi) = \phi$, therefore, 
  the corresponding boundary condition simply reads $|\df/\phi| < \infty$.
  
  We can also notice that, in all cases under consideration, the boundary conditions 
  at flat spatial infinity are the same as for Fisher's solution, that is, $|Y| < \infty$ as 
  $z\to\infty$.    
  
  In BD theory with the field \rf{phi-BD}, the static solution reads
\beq             \label{ds-BD}
  		ds^2_J = P^{-\xi} \Big[ P^a dt^2 - P^{-a} dx^2 - x^2 P^{1-a} d\Omega^2 \Big], 
  	 \quad\  		\phi = P^{\xi}, \qq 
   		P:= 1 -\frac{2k}{x}, \quad\ \xi :=\pm \sqrt{\frac{1-a^2}{2\omega+3}}.
\eeq  
   The solution is defined, just as Fisher's, in the range $x \in (2k, \infty)$, and there is a 
   naked singularity at $x = 2k$, although in some region of the parameter space 
   the spherical radius (such that $r^2_J = -g^{(J)}_{\theta\theta} = x^2 P^{1-a-\xi}$)
   blows up as $x\to 2k$. A more detailed description of this solution in the same notations
   can be found in \cite{skvo10}. The only exception is the case $\omega =0, \ a =\xi =1/2$,
   when we obtain the so-called ``\Scw\ \wh'' with the metric
\beq                \label{BD-wh}
  		  ds_J^2 = dt^2 - (1-2k/x)^{-1} - x^2 d\Omega^2, 	
\eeq  
  where $x = 2k$ is a throat beyond which there is one more region with the same metric.
  
  Due to \rf {phi-BD}, for perturbations near $x = 2k$ we have $\df \sim \phi \dpsi$,
  and the boundary condition $|\df/\phi| < \infty$ is equivalent to $\dpsi < \infty$, a
  condition more stringent than $|\dpsi/\psi| < \infty$ that we have used for Fisher's 
  solution. As a result, in the expression \rf{Y-2k} we should require $C_2 =0$,
  and we obtain a complete boundary-value problem for \eqn{Schr} with the boundary
  conditions $|Y|< \infty$ as $z\to \infty$ and $Y\sim \sqrt{z}$ as $z \to 0$. 
  
\paragraph{Solution of Barker's theory.} In Barker's theory \rf{Bark}, according to 
  \rf{Bark-b}, we have $ds_J^2 = \cos^2 (\psi-\psi_0)\ ds_E^2$, where $ds_E^2$
  is given in \rf{fish-a} with the scalar field $\psi \sim \log (1 - {2k}/x)$.
  Assuming $\psi_0$ within the interval $(-\pi/2, \pi/2)$, the solution is \asflat\
  at large $x$. As $x$ decreases, the field $\psi$ inevitably reaches a value $x_s$ at 
  which $\cos (\psi-\psi_0) =0$, and there the metric  $ds_J^2$ has a curvature 
  singularity. Thus the solution exists in the range $x_s < x < \infty$. 
  
  If we require that the perturbation $\df$ of the field $\phi =1/\cos^2 (\psi-\psi_0)$
  should satisfy the condition  $|\df/\phi| < \infty$, then, as is easily verified,
  for the corresponding $\dpsi$ near the singularity $x = x_s$ we must require 
  $\dpsi = O(x-x_s)$. The radius $r$ is finite at $x=x_s$, therefore, the same condition 
  is valid for $Y(z)$:  $Y = O(x-x_s)$. 
  
  Thus the stability analysis reduces to a boundary-value problem for \eqn{Schr} on the 
  interval $(x_s, \infty)$ with the conditions $|Y|\to 0$ as $x\to \infty$ and 
  $|Y|/(x-x_s) < \infty$ as $x \to x_s$. It is also clear that at such an intermediate
  point the tortoise coordinate $z$ smoothly depends on $x$ according to \rf{u_to_z},
  therefore the same condition may be rewritten as $|Y|/(z-z_s) < \infty$ as $z \to z_s$,
  where $z_s = z(x_s)$.  
  
  We still notice that the effective potential $\Veff$ is known in terms of $x$ 
  rather than $z$, while the boundary conditions are equally well formulated using 
  $z$ or $x$. It therefore seems that the problem, if solved numerically, can be better 
  treated using the coordinate $x$, or with the harmonic coordinate $u$ used in 
  \rf{fish}. Note that in these coordinates the master equation does not have its
  canonical form \rf{Schr} but, instead, has the form \rf{eq-X} in which all 
  coefficients are known explicitly.  
    
\paragraph{Solution of Schwinger's theory.}  We now have 
  $ds_J^2 = [(\psi-\psi_0)^2/K] ds_E^2$, and $\phi = K/(\psi-\psi_0)^2$.
  The solution is \asflat\ at large $x$ under the condition $\psi_0 \ne 0$, 
  and splits into two branches, $\psi_0 >0$ and  $\psi_0 < 0$. 
  
  In the case $\psi_0 > 0$, the solution is defined for $x \in (x_s , \infty)$,
  and the singularity at $x = x_s > 2k$ is quite similar to that in Barker's theory, 
  therefore, we have again the boundary condition  $|Y|/(x-x_s) < \infty$ as $x \to x_s$.
  
  In the case $\psi_0 < 0$,  the solution is defined in the whole range $(2k , \infty)$,  
  and the boundary condition is $|\df/\phi| \sim |\dpsi/\psi| < \infty$, quite
  similar to the one for Fisher's solution. We thus come to the same conclusion, that 
  this solution is unstable under radial perturbations. 
  
\paragraph{Scalar fields with nonminimal coupling to gravity, $f(\phi) = 1 - \xi\phi^2$.} The 
  analytical form of the solutions in the theories \rf{nonmin} and the properties of the 
  corresponding geometries depend on the nonminimal coupling constant $\xi$. The following 
  four cases are distinguished:
\begin{description}   
\item[(i)]
	$\xi = 1/6$, conformal coupling.
	From \rf{phi-psi} we obtain, assuming $\phi^2 < 6$ (to provide $f >0$),  
\beq
		\psi-\psi_0 = \frac{\sqrt 3}{2} \log \frac{\sqrt 6 + \phi}{\sqrt 6  \phi}
		\quad \then \quad 
		\phi = \sqrt 6 \tanh \frac{\psi-\psi_0}{\sqrt 3}.
\eeq	
     Accordingly, 
\beq
		 ds^2_J = \cosh^2 \frac{\psi-\psi_0}{\sqrt 3} ds_E^2.	
\eeq     
    The Jordan-frame metric behavior near $x=2k$ ($\psi \to \infty$) depends on the parameter $a$:\\
    if $a < 1/2$, we have an attracting singularity with $g_{00}^J \to 0$ and 
    $r^2_J = -g^J_{22} \to \infty$; \\
    if $a > 1/2$, there is a repulsive singularity with $g_{00}^J \to \infty$ and 
    $r^2_J = -g^J_{22} \to 0$; \\
    Lastly, if $a=1/2$, the sphere $x=2k$ is regular, and a conformal continuation further leads 
    either to a singular center ($r_J \to 0$), or to a wormhole, or to an extremal black hole. This 
    behavior has been studied in detail in many papers, see, e.g,, \cite{bbm70, kb73, br-book, Turok92}, 
    including, in particular, stability studies. In the present paper we assume $a \ne 1/2$, so that
    the solution is defined for $x > 2k$, and $x=2k$ with $\psi = \infty$ is a singularity.     
    
    At large $\psi$ we obtain
\beq
            f  = \bigg(\cosh\frac{\psi-\psi_0}{\sqrt 3}\bigg)^{-2} \sim \e^{- 2\psi/\sqrt 3}.
\eeq            
    Then, for perturbations, similarly to the Brans-Dicke theory, we have $\delta f \sim f\dpsi$,
    and the boundary condition at $x=2k$ again reads $|\dpsi| < \infty$.     
	
\item[(ii)]
	$0 < \xi < 1/6$. According to \rf{phi-psi}, the fields $\phi$ and $\psi$ are related by
\beq                        \label{xi-2} 
		\sqrt 2 d\psi = \frac {\sqrt{1 - \eta \phi^2}}{1 -\xi\phi^2}d\phi,
		\qq
		\eta = \xi(1 - 6\xi); \qq \eta >0.
\eeq	
	Its integration gives a result that can be written, following \cite{bar-vis2, stepan1}, as
\bearr                       \label{psi-2} 
		\psi - \psi_0 = \frac{\sqrt 3}{2} \log \Big[B(\phi) H^2(\phi)\Big],
\\ \lal                       \label{B-2} 
		B(\phi) = \frac{\sqrt{1-\eta\phi^2} + \sqrt 6 \xi\phi}{\sqrt{1-\eta\phi^2} - \sqrt 6 \xi\phi},
\\ \lal                       \label{H-2} 
           \log H(\phi) = \frac{\sqrt\eta}{\sqrt 6 \xi} \arcsin(\sqrt\eta \phi).
\ear			 
     It is easy to see that $\phi \to 1/\sqrt\xi$ leads to $\psi \to \infty$ which in turn corresponds to
     $x \to 2k$. The function $H(\phi)$ is finite and smooth in the whole range $\phi^2 < 1/\xi$,
     therefore, the qualitative properties of the solution in \MJ\ are completely determined by 
     $B(\phi)$, and the results are quite similar to those for $\xi = 1/6$. In particular, 
     the solution with $a = 1/2$ requires a conformal continuation and is excluded from the present 
     analysis. At $a \ne 1/2$, at large $\psi$ the function $f(\phi(\psi))$ behaves as 
     $\e^{-2\psi/\sqrt 3}$, and for perturbations we obtain the same boundary condition at 
     $x = 2k$: $|\dpsi| < \infty$. 
	
\item[(iii)]
	$\xi > 1/6$. In this case, \eqs \rf{xi-2}--\rf{B-2} are again valid, except that now  $\eta < 0$.
	This leads to another expression for $H(\phi)$:
\beq                       \label{H-3} 
			\log H(\phi) = -\frac{\sqrt{-\eta}}{\sqrt 6 \xi} \sinh^{-1} (\sqrt{-\eta} \phi).
\eeq	 
     Still, just as with $\xi < 1/6$, the function $H(\phi)$ is finite and smooth at $\phi^2 < 1/\xi$,
     hence all further inferences simply repeat those for $0 <\xi <1/6$.  

\item[(iv)]
	$\xi < 0$. 			
	Equations \rf{xi-2}--\rf{B-2} are again valid, while now both $\xi < 0$ and $\eta < 0$. Instead of 
	\rf{H-3}, we now obtain 
\beq                       \label{H-4} 
			\log H(\phi) = \frac{\sqrt{-\eta}}{\sqrt 6 |\xi|} \sinh^{-1} (\sqrt{-\eta} \phi).
\eeq	
     Unlike the previous cases, the $\phi$ field is now defined for all $\phi \in \R$, Still now the 
     function $B(\phi)$ is finite in the whole range of $\phi$, and the solution properties are 
     governed by $H(\phi)$. At large $\psi$ we have $f(\phi(\psi)) \sim \e^{2\psi/\sqrt 3}$, the
     conformal factor $1/f$ in \rf{map} tends to zero and only strengthens the attracting 
     singularity of Fisher's solution, and no conformal continuations are observed now at any $a$.
     Furthermore, with this asymptotic behavior of  $f(\phi(\psi))$ at large $\psi$, we again obtain 
     the boundary condition for perturbations in the form $|\dpsi| < \infty$ at $x=2k$.
\end{description}  
  
\subsection{Boundary-value problems}

  We see that, in addition to the cases where the final instability conclusion was achieved 
  without solving the master equation, we have two boundary-value problems to be solved
  numerically:  
\begin{enumerate}
\item[\bf 1.]
		Equation \rf{Schr}, range $z\in (z_s > 0,\infty)$, boundary conditions: 
		$|Y| < \infty$ as $z \to \infty$, and $|Y|/(z -z_s) < \infty$ as $z \to z_s$.   	
\item[\bf 2.]
		Equation \rf{Schr}, range $z\in (0,\infty)$, boundary conditions: 
		$|Y| <\infty$ as $z \to \infty$, and $|Y|/\sqrt{z} < \infty$ as $z \to 0$. 
\end{enumerate}  
  
   The purpose of solving these problems is to find out whether there are physically 
   meaningful solutions (i.e., those satisfying the above boundary conditions) 
   with eigenvalues $\Omega^2 \leq 0$, which would mean that the background 
   static solutions are unstable. 
  
   In both cases, the boundary conditions do not coincide with those used in quantum 
   mechanics for the one-dimensional Schr\"odinger equation \rf{Schr} (where the energy
   level $E$ would appear instead of $\Omega^2$): indeed, in QM we would require 
   the normalization $\int Y^2(z) dz =1$, hence quadratic integrability of $Y(z)$.
    
   Though, in our problems, since $\Veff \to 0$ as $z\to \infty$, our requirement of finite $Y$ 
   leads to the same conclusion for the ``energy levels'' $\Omega^2 < 0$ or $ E < 0$:
   we have, at large $z$, $Y \approx C_1 \e^{|\Omega|z} + C_2 \e^{-|\Omega|z}$,
   and in both cases we have to require $C_2 =0 \ \then\ Y \to 0$ as $z \to \infty$.
   On the other hand, the assumption $\Omega =0$ leads to $Y \approx C_1 + C_2 z$, 
   then in our problems it is sufficient to put $C_2 =0$ while a nonzero $C_1$ is admitted,
   whereas quadratic integrability would require $C_1=C_2 =0$. (Note that with 
   $\Omega =0$ the perturbation $\dpsi$ or $\df$ can grow linearly with time.)
   
   From the form of the potential \rf{Veff-x}, since $\Veff \approx  2ak/z^3 > 0$ at large 
   $z\approx x$, we know that $\Veff > 0$ in some range $z > z_0$, where $z_0$ depends
   on $a$ and $k$ (see Fig.\,1). For Problem 1 it directly follows that there are no 
   eigenvalues $\Omega^2 \leq 0$ if the ``left boundary'' $z_s \geq z_0$. Indeed, 
   suppose that $Y \to Y_1 > 0$ as $z \to \infty$. Then, with $\Omega^2 \leq 0$,
   and $\Veff > 0$, from \eqn{Schr} we obtain $d^2 Y/dz^2 > 0$, hence, as $z$ decreases
   from infinity, $Y$ can only grow, remaining everywhere positive, and is unable to reach 
   zero at $z=z_s$, as required by the second boundary condition. Quite similarly, if 
   $Y \to Y_1 < 0$ as $z \to \infty$, we obtain everywhere $d^2 Y/dz^2 < 0$, and again 
   $Y(z_s) =0$ cannot be achieved. The same reasoning works as well if $Y(\infty) = 0$, 
   since a nontrivial solution to \eqn{Schr} must deflect from zero either to $Y>0$, which 
   leads to $d^2 Y/dz^2 > 0$, or to $Y <0$ leading to $d^2 Y/dz^2 < 0$, again making 
   impossible $Y(z_s)=0$. Even more than that, it is clear that if we follow the function 
   $Y(z) >0$ (say) to values smaller than $z_0$, at which $\Veff < 0$, this function,
   being smooth enough, cannot immediately reach zero --- hence it follows that 
   an eigenvalue $\Omega^2 \leq 0$ can only be obtained at $z_s \leq z_{\rm crit} < z_0$,
   and such a critical value $z=z_{\rm crit}$ (or its corresponding value of $x$)
   can probably be found only numerically.
   
   Unlike that, Problem 2 deals with the full range $z > 0$ ($x > 2k$), it does not contain 
   such a free parameter as $z_s$ and should be considered directly.
   
\section{Numerical analysis}  
   
   In both boundary-value problems 1 and 2, for a numerical analysis it is better to
   choose the quasiglobal independent variable $x$ since both the static solution and 
   the perturbation equations involve functions of $x$ whereas the relation \rf{u_to_z}
   after integration leads to $z(x)$ in terms of the hypergeometric function, making it
   impossible to express all quantities in terms of $z$. Thus let us use \eqn{eq-X} 
   that has the following form in terms of $x$ for the metric \rf{fish-a}:
\bearr                 \label{X''}
		X'' + \frac{2 (x-k)}{x (x- 2k)} X'
			+ \bigg[\frac{x^{2a}}{(x-2k)^{2a}}\Omega^2 - W(x)\bigg] X =0,	
\yyy				\label{W(x)}
		W(x) =  - \frac{2k^2 (1 - a^2)}{x (x -2k) (x - k - ak)^2}.
\ear 
   
   Then, with the relationships \rf{X_to_Y} and \rf{u_to_z}, or equivalently, since
   $|X| = |\dpsi|$, the two boundary-value problems to be solved are reformulated 
   as follows:
\begin{enumerate}
\item[\bf 1.]
		Equation \rf{X''}, range $x\in (x_s, \infty)$ ($x_s > 2k$), boundary conditions: 
		$X \to 0$ as $x \to \infty$, and $|X|/(x - x_s) < \infty$ as $x \to x_s$.   	
\item[\bf 2.]
		Equation \rf{X''}, range $x\in (2k,\infty)$, boundary conditions: 
		$X \to 0$ as $x \to \infty$, and $|X| < \infty$ as $x \to 2k$. 
\end{enumerate}  
   In both problems, we seek values of $\Omega^2 < 0$, such that solutions to
   the master equation satisfy these boundary conditions. If there are such 
   $\Omega^2$, the corresponding static solutions of STT are unstable.
   
   In the case where perturbations with $\Omega =0$ are meaningful, which 
   also leads to an instability, the approximate general solution to \eqn{X''} at large $x$
   reads $X = C_1 + C_2/x$ (in agreement with $Y(z) = C_1 z + C_2$, since 
   $z \approx x$ and $Y \approx xX$), and to have $X \to 0$ as $x \to \infty$ 
   in both Problems 1 and 2, we should simply require $C_1 =0$.  
   
   So the boundary-value problems have been posed, but their numerical solution 
   is complicated by the infinitely remote ``right end'' $x \to \infty$, and for their practical 
   solution it makes sense to move the boundary to some large but finite value $x_0$.
   Then, to solve the equation, we should specify the values of $X(x_0)$ and $X'(x_0)$
   corresponding to an admissible solution. With $\Omega^2 = - K^2,\ K >0$, such 
   a solution has the asymptotic form 
\beq                  \label{X'as}   
   		X(x) \approx \frac 1x \e^{-Kx}  \ \then \  
   		\frac{X'}{X} \approx - \frac 1{x_0} - K.
\eeq
   Therefore, taking an arbitrary value of $X_0 = X(x_0)$ (its particular value does not
   matter since the equation to be solved is linear), we must specify the derivative 
   $X'(x_0) = - X_0 (K + 1/x_0)$. Then, solving \eqn{X''} with such boundary conditions 
   and different $K$, we will try to select such $K = |\Omega|$ that the boundary condition 
   on the ``left end'' is satisfied (the ``shooting'' method).  
   
   In the case $\Omega =0$, it is easy to determine that an admissible solution has 
   the asymptotic form $X \approx C_2/x$ at large $x$ with an arbitrary constant
   $C_2$, hence the derivative is $X' \approx - C_2/x^2$, and the conditions
   at large $x_0$ should be specified taking into account these asymptotic expressions.
   
   On equal grounds, the ``shooting'' may also be implemented ``from left to right'', 
   that is, by specifying the appropriate boundary conditions at $x = x_s$ in Problem 1 
   and at $x = 2k$ in Problem 2. However, in the latter case, we have to step aside from 
   $x =2k$ because $x-2k$ appears in the denominators in \rf{X''}
   
   Unlike that, in Problem 1 we can safely use the conditions $X =0$, $X' = X'_0 \ne 0$ 
   at $x = x_s$ and seek solutions well-behaved at infinity. In doing so, it makes sense 
   to consider only sufficiently small $x_s$, at which the potential \rf{Veff-x} $\Veff < 0$, 
   otherwise, quite evidently, eigenvalues $\Omega^2 \leq 0$ are absent due to 
   the reasoning at the end of the previous section.   
   
\subsection{Boundary-value problem 1}

  To solve \eq\eqref{X''}, we use the shooting method described above for various values of $a$, $x_s$,
   and $\Omega^2$.  Without loss of generality we put $k=1$ (thus fixing the length scale) and the 
   initial value of the derivative, $X'(0)=1$. The standard procedure of ODE solving leads to a numerical 
   curve $X_\text{num}(x)$ corresponding to a chosen test value of $\Omega^2<0$. This curve has 
   a critical behavior with a drastically fluctuating sign at large $x = x_r\gtrsim 1000$ at any non-eigenvalue 
   $\Omega^2$. Tracking the value of $X(x_r)$ allows us to ``catch'' the critical value of $\Omega^2$ 
   as a candidate for an eigenvalue with necessary accuracy. The result of such a numerical analysis is 
   depicted in Fig.\,2 for three values of $a$. Each curve in this plot is actually a separatrix that 
   separates regions of different signs of $X_\text{num}(x)$ at large $x_r$ during the shooting procedure. 
   The pairs $(\Omega^2 < 0,x_s)$ along this separatrix, where $X_\text{num}(x_r)\cong 0$ up to a 
   numerical error, correspond to solutions of the eigenvalue problem at fixed $a$ and indicate 
   an instability of the system under consideration.

   As is evident from Fig.\,2 (left panel), the instability is only observed for $x_s$ inside quite a small 
   interval $(2k, \tilde x_s)$, where $k=1$. When $x_s\to \tilde x_s$, the corresponding eigenvalue 
   $\Omega^2\to -0$. For each $a$, the upper boundary $\tilde x_s$ can be found by numerically 
   solving \eq\eqref{X''} with $\Omega:=0$. In this case we obtain a curve $X_\text{num}(x)$ that tends 
   to some $C_1=\const$ at large $x$. This $C_1$ depends on $x_s$ and vanishes at some $x_s=\tilde x_s$ 
   (see Fig.\,2, right panel). For $a=0.4,\ 0.6,\ 0.8$ one can numerically find 
   $\tilde x_s \simeq 2.0018,\ 2.0090,\ 2.0163$, respectively. These values of $x_s$ indicate 
   the existence of a solution $X(x)$ corresponding to the eigenvalue $\Omega=0$.

   The eigenfunctions $X(x)$ for different $a$ and $x_s$ are plotted in Fig.\,3.
\begin{figure*}[t]
\centering
\includegraphics[width=8cm]{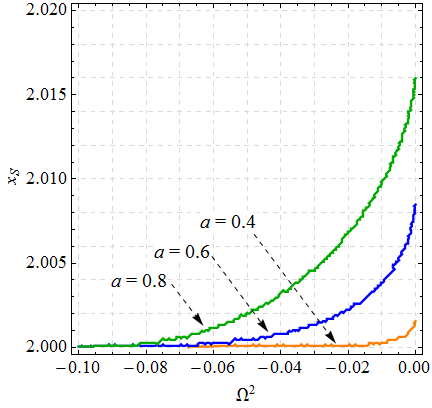}\qq
\includegraphics[width=8.5cm]{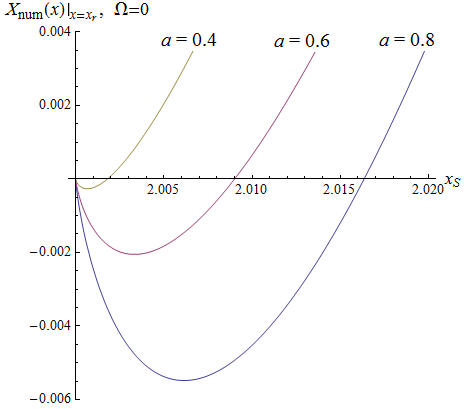}\qq
\caption{\small
		Boundary-value problem 1. \underline{Left}: A diagram showing the existence of possible 
		eigenvalues $\Omega^2 < 0$ depending on the choice of $x_s$ for $a=0.4,\ 0.6,\ 0.8$. 
		Each curve separates regions of $X_\text{num}(x) > 0$ at large $x= x_r$ (above the line)
		from regions of $X_\text{num}(x) < 0$ (below the line). \underline{Right}: The behavior
		of the asymptotic value $X_\text{num}(x_r) = C_1$ as a function of $x_s$ in solutions with 
		$\Omega=0$ for $a=0.4,\ 0.6,\ 0.8$. The zeros of the corresponding curves coincide with 
		the upper bounds $\tilde x_s$ of the instability intervals.}
\end{figure*}  
\begin{figure*}[ht]
\centering
\includegraphics[width=8.5cm]{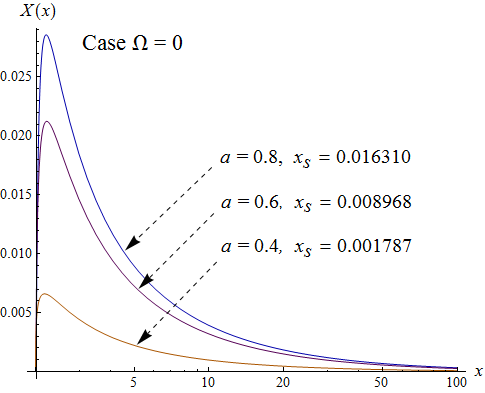}\qq
\includegraphics[width=8.5cm]{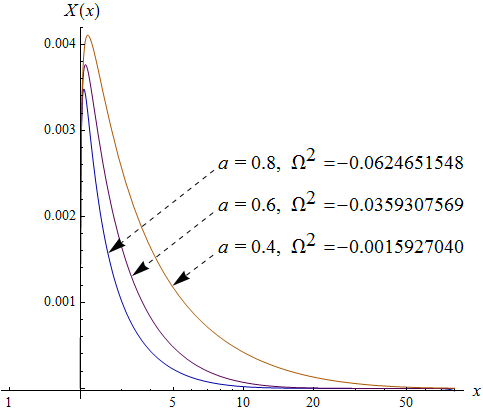}\qq
\caption{\small
		Boundary-value problem 1. \underline{Left}: Solutions $X(x)$ with $\Omega=0$ for 
		$a=0.4,\ 0.6,\ 0.8$. The corresponding values of $x_s$ are presented in the plot. 
		\underline{Right}: Solutions $X(x)$ for $x_s=2.001$ and $a=0.4,\ 0.6,\ 0.8$. The 
		corresponding eigenvalues $\Omega^2$ are presented.}
\end{figure*}

\subsection{Boundary-value problem 2}
   
  An attempt to numerically solve \eqn{X''} ``from right to left,'' specifying the boundary 
  condition at large $x$ using \eqn{X'as} leads to a numerical instability near the singularity $x =2k$. 
  Therefore, it makes sense to launch the ``left to right'' procedure, specifying the boundary
  condition close to $x = 2k$, where a desired solution to \eqn{X''} should tend to a constant 
  value. In doing so, it is not possible to require simply $X = X_0$ and $X' =0$ at some $x_0$
  close to $2k$ since quite evidently the last term in \rf{X''} proportional to $X$ 
  blows up as $x\to 2k$ and must create if not large then nonzero values of the derivative $X'$.
  To find suitable boundary conditions, let us find the asymptotic behavior of $X(x)$ near $x=2k$ 
  in a better approximation than $X =\const$.
  
  For convenience, let us rewrite \eqn{X''} in terms of $y = x-2k$,
\bearr                 \label{X''y}
		X'' + \frac{2 (y + k)}{y (y + 2k)} X'
			+ \bigg[\frac{(2k + y)^{2a}}{y^{2a}}\Omega^2 - W(y)\bigg] X =0,	
\yyy				\label{W(x)}
		W(y) =  - \frac{2k^2 (1 - a^2)}{y (2k+y) (k - ak+y)^2},
\ear 
  and seek solutions at small $y$ in the form 
\beq  			\label{X-p}
			X(y) = X_0 + c y^p, \qq  X_0, \ c,\ p = \const, \qq p > 0.
\eeq  
  The solutions are different for different values of $a$:
  
\begin{enumerate}  \itemsep 1pt
\item $a < 1/2$. 
	In this case, the term with $\Omega^2$ in \eqn{X''y} grows slower 
	than $W(y) \sim 1/y$ as $y \to 0$ and can be neglected. Then, substituting 
	\rf{X-p} to \rf{X''y} and preserving the dominant order of  magnitude, we obtain 
\[
		cp^2 y^{p-2}+ \frac{(1+a) X_0}{y} =0 \ \then \ p =1, \ \ c = -(1+a) X_0.
\]	
     In this approximation, the boundary conditions at $x = 2k+y_0$, $y_0\ll 1$, read
\beq
		X= X_0[1 -(1+a) y_0], \qq X' = -X_0(1+a).
\eeq      		  
\item $a = 1/2$. 
     Then both terms with $\Omega^2$ and $W(y)$ behave as $1/y$,
	and similarly to the previous case we obtain the constants
\[
		p=1, \ \ \  c = -X_0 (2k \Omega^2 + 3/2),
\]	        
    and the boundary conditions at $x = 2k+y_0$, $y_0\ll 1$ in the form
\beq
		X = X_0[1 - (2k \Omega^2 + 3/2)y_0], \qq X' = -X_0(2k \Omega^2 + 3/2).
\eeq    
\item $a > 1/2$. 
	Now the term with $\Omega^2$ in \eqn{X''y} grows faster than $W(y) \sim 1/y$ 
	as $y \to 0$, and the latter can be neglected. Then in the same manner as before
	we obtain the constants
\[  
		p = 2(1-a) < 1, \qq   c = - X_0 \Omega^2 \frac{(2k)^{2a}}{p^2},
\]
  and the boundary conditions at $x = 2k+y_0$, $y_0\ll 1$
\beq
		X = X_0\bigg[1 - \Omega^2 \frac{(2k)^{2a}}{p^2}y_0^p\bigg], \qq 
		X' = - X_0 \Omega^2 \frac{(2k)^{2a}}{p} y_0^{p-1}.
\eeq    
   One can notice that under these conditions the derivative $X'$ is quite large at small $y_0$,
   like that of the function $\sqrt{x}$ near $x=0$.
\end{enumerate}

  We implement the shooting method separately for each of the above three cases, putting $k=1$ 
  and $X_0=1$ without loss of generality. We also put $y_0=10^{-5}$ for suitable accuracy of 
  the numerical plots. It is of interest to study the dependence of possible eigenvalues $\Omega^2$ 
  (if any) on the parameter $a$. Using the method of tracking fluctuations of $\sign X(x)$ at large 
  ``right end'' (as in the previous section), one can draw a corresponding separatrix, along which 
  the pairs $(a, \Omega^2)$ provide a solution to the eigenvalue problem.

  The results of this analysis are depicted in Fig.\,4 (left panel). One can see that despite the different 
  nature of boundary conditions in cases 1--3 above, the resulting separatrix is continuous, including 
  the particular point $(a=1/2,\ \Omega^2\simeq -0.1918931029)$ corresponding to the special case 
  $a = 1/2$. This diagram allow us to draw a conclusion (at least up to the numerical accuracy governed 
  by the choice of a small but finite starting value $y_0$) that there is an instability for all $a \in (0, 1)$
  due to the existence of the eigenvalues $\Omega^2 < 0$. Examples of numerical solutions 
  $X_\text{num}(x)$ are shown in the right panel of Fig.\,4.
\begin{figure*}[ht]
\centering
\includegraphics[width=7.2cm]{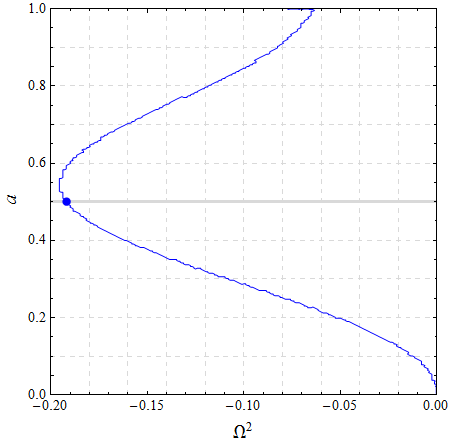}\qq
\includegraphics[width=8.5cm]{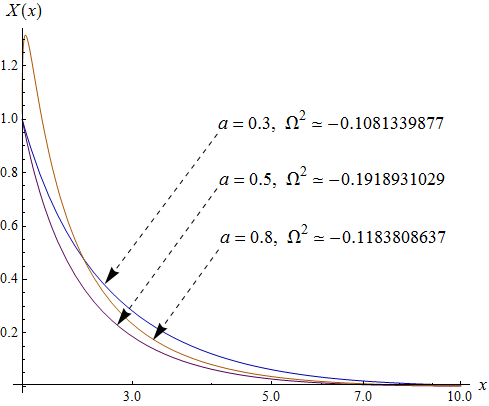}
\caption{\small
		 Boundary-value problem 2. \underline{Left}: a numerical diagram showing the existence of 
		 eigenvalues $\Omega^2$ depending on the chosen value of $a$. The curve is a separatrix 
		 between regions of different signs of $X_\text{num}(x_r)$ at large ``right end'' $x_r\gtrsim 1000$ 
		 during the shooting procedure with $y_0=10^{-5}$. \underline{Right}: numerically found
		 eigenfunctions $X(x)$ for $a=0.3,\ 0.5,\ 0.8$ with the corresponding eigenvalues $\Omega^2$.}
\end{figure*}
	
	A note on the obtained increment values $|\Omega|$ of unstable perturbations. In Fisher's solution 
	\rf{fish-a}, the \Scw\ mass is $M = ak$. The conformal factor $1/f$ changes this value, but generically
	does not change its order of magnitude, therefore, since $a$ is of the order of unity, we can say that 
	in our STT solutions we also have $M \sim k$. Therefore, with our scale fixing $k=1$, the obtained 
	values $|\Omega| \lesssim 1$ mean that the characteristic times of perturbation growth are 
	approximately of the order of or slightly larger than the time needed for a photon to cover a 
	distance $\sim M$, the \grav\ radius of the corresponding \Scw\ mass.

\section{Conclusion}    

   We have considered the linear stability problem for conformally Fisher space-times that comprise 
   vacuum solutions of the Bergmann-Wagoner-Nordtvedt class of STT of gravity with 
   massless scalar fields. Since the perturbation equations for these space-times, as a 
   special case of \grav-scalar field equations, formulated in the Jordan frame manifold \MJ, 
   are transformed to the corresponding equations in \ME, the latter, being common for all STT 
   of the present class and coinciding with those of GR, acquire a universal meaning. Unlike that, 
   the boundary conditions that select physically meaningful perturbations, depend on a particular
   STT and even on the properties of particular solution. 
   
   In some cases, the boundary conditions turn out to coincide with those already known in
   \ME --- in such cases the stability conclusions for \MJ\ also coincide with those for \ME. 
   In other cases, it is necessary to consider separate boundary value problems. This work has 
   been done here for four well-know examples of STT, and its results are summarized in Table 1.   
   
\begin{table*}
\caption {Massless STT solutions: Stability under monopole perturbations}
\begin{center}
\small
\begin{tabular}{|p{65mm}|p{70mm}| l |}
\hline
	\cm Theory, solution           & \cm Description${}^*$   & Results$^{**}$   \tall
\\[2pt]  \hline  \tall
		Brans-Dicke, $\omega\ne 0$ &   
								Singularity at $x=2k$, BVP 2   & unstable
\\[2pt]
		Brans-Dicke, $\omega=0$, $a \ne 1/2$  &
								Singularity at $x=2k$, BVP 2   & unstable
\\[2pt] \hline  \tall
		Barker, $f = 1/\cos^2(\psi-\psi_0)$  &
						          Singularity at $x=x_s > \tilde x_{s}(a) $, BVP 1 & stable
\\[2pt] 
		Barker, $f = 1/\cos^2(\psi-\psi_0)$  &
						          Singularity at $x=x_s \leq \tilde x_{s}(a) $, BVP 1 & unstable
\\[2pt] \hline  \tall
		Schwinger, $f =(\psi-\psi_0)^2/K$, $\psi_0 >0$   &
								Singularity at $x=x_s > \tilde x_{s}(a) $, BVP 1 & stable
\\[2pt] 
		Schwinger, $f =(\psi-\psi_0)^2/K$, $\psi_0 >0$   &
								Singularity at $x=x_s \leq \tilde x_{s}(a) $, BVP 1 & unstable
\\[2pt] 
		Schwinger, $f =(\psi-\psi_0)^2/K$, $\psi_0 < 0$   &
								Singularity at $x=2k$ similar to Fisher's    	& unstable	+
\\[2pt] \hline    \tall
		Nonminimal coupling, $\xi >0$, $a\ne 1/2$  &
								Singularity at $x=2k$, BVP 2   & unstable
\\[2pt]
		Nonminimal coupling, $\xi < 0$,  &
								Singularity at $x=2k$, BVP 2   & unstable
\\[2pt] \hline
\end{tabular}    
\end{center}
\small
   	${}^*$ BVP = boundary-value problem.\\
   	${}^{**}$ ``unstable'' means that there is a perturbation mode with a finite increment $|\Omega|$;
   		``unstable+'' means that a perturbation increment is indefinite, as it is for Fisher's solution.
\end{table*}                
   
   It is clear that the present results are applicable to many more particular STT because the 
   boundary-value problems 1 and 2 for spherical perturbations emerge under quite general 
   conditions. In particular, it is true for various STT representations of various
   modified theories of gravity other than STT as such, for example, the original \cite{hybrid}
   and generalized \cite{g-hybrid} versions of hybrid metric-Palatini gravity, for which \ssph\ solutions
   are discussed in \cite{we-hyb, we-g-hyb}, as well as different models of nonlocal and 
   high-order gravity, see, e.g., \cite{od17-rev, od08, kb-eliz10, chrv22}.
   
   Let us also recall that the present results have been obtained only for STT with massless scalars, 
   while there are numerous models and solutions with nonzero scalar field potentials $U(\phi)$,
   see \eqn{S_J}; note that $f(R)$ theories of gravity are also reducible to STT with nonzero $U(\phi)$. 
   Moreover, we have restricted ourselves to canonical (nonphantom) STT, whereas models
   with phantom scalars are also of much interest and even more diverse. In all cases, except for 
   those with conformal continuations, it is possible to reduce the perturbation equations to 
   those in the Einstein frame, but the appropriate boundary conditions must be formulated 
   ``individually'' for each separate solution.  

\Funding
{The research of K. Bronnikov, S. Bolokhov and M. Skvortsova was supported by 
	RUDN University Project FSSF-2023-0003.
  K. Bronnikov was also supported by the Ministry of Science and Higher Education 
  of the Russian Federation Project No. 0723-2020-0041.	
  K. Badalov and  R. Ibadov  gratefully acknowledge the support from Ministry of Innovative 
  Development of the Republic of Uzbekistan, Project  No. FZ-20200929385.     
}   
        

\end{document}